\documentclass[aps,pra,twocolumn,groupedaddress,amsmath,amssymb]{revtex4-1}
\usepackage{mathtools}
\usepackage{graphicx,caption,subcaption}  
\usepackage{dcolumn}   
\usepackage{bm}        
\usepackage{verbatim}   
\usepackage{wrapfig}
\usepackage{float, array}
\usepackage{amsmath}
 \usepackage{amssymb}
\usepackage{verbatim, url}
\usepackage{fullpage}
\usepackage{enumerate}
\usepackage{bm}
\usepackage{notes2bib}
\usepackage{lpic}
\usepackage{pgfplots}
\usepackage[T1]{fontenc}
\usepackage[latin1]{inputenc}
\usepackage[extra]{tipa}
\usepackage{accents}
\usepackage{epstopdf}
\usepackage[titletoc,title]{appendix}
 \usepackage{amsthm}

\newlength{\dhatheight}

\newcommand{\bra}[1]{\langle #1|}
\newcommand{\ket}[1]{|#1\rangle}

\newcommand{\nn}{\nonumber}
\newcommand{\f}[2] {\frac{#1}{#2}}
\newcommand{\p}{\partial}
\newcommand{\m}[1]{\mathcal{#1}}

\newcommand{\beq}{\begin{equation}}
\newcommand{\eeq}{\end{equation}}
\newcommand{\beqn}{\begin{eqnarray}}
\newcommand{\eeqn}{\end{eqnarray}}
\newcommand{\la}{\left\langle}
\newcommand{\ra}{\right\rangle}

\DeclarePairedDelimiter\abs{\lvert}{\rvert}%
\DeclarePairedDelimiter\norm{\lVert}{\rVert}%

\makeatletter
\let\oldabs\abs
\def\abs{\@ifstar{\oldabs}{\oldabs*}}
\let\oldnorm\norm
\def\norm{\@ifstar{\oldnorm}{\oldnorm*}}
\makeatother

\begin{document}

\title{Classical simulation of arbitrary quantum noise}
\author{Seyyed M.H. Halataei}
\affiliation{
   Department of Physics, University of Illinois at Urbana-Champaign,\\
   1110 West Green Street, Urbana, Illinois 61801, USA
   }
\date{Oct. 24, 2017}
\begin{abstract}
I present an explicit classical simulation of arbitrary quantum noise for quantum models in which one qubit interacts with a quantum bath. The classical model simulates the interaction of the bath and the qubit by random unitary evolutions. I show that any arbitrary quantum dynamics, including quantum dissipation, recurrence, and dephasing, can be simulated classically when one allows the unitary operators in the classical model to depend on the initial state of the system and bath. For initial mixed states of the system and non-product states of the system and bath, I demonstrate that random unitary expansion is still possible, in terms of a set of pure states.
  \end{abstract}

\maketitle

\section{Introduction}
Entanglement is ``\emph{the} trait of quantum mechanics'' that ``enforces its entire departure from classical lines of thought'', according to  Schr\"{o}dinger \cite{Schrodinger1935}. In open quantum
systems, the creation of entanglement between the system and its environment is commonly stated as the root cause of decoherence \cite{Blume-Kohout2008, Bellomo2007, Kubotani2008, Gorin2007, Rivas2014, Aolita2015, Franco2013, deVega2017, Mortezapour2017, Gonzalez2016}. The loss of coherence in open quantum systems, or decoherence, itself is considered to be the reason for the appearance of classical traits in quantum systems and to be connected with the quantum-to-classical transition \cite{Zurek2003,Joos2003}. 

However, recently it has been demonstrated that certain types of decoherence can be simulated classically by random unitary dynamics without appealing to entanglement with an environment and the idea of transformation of information, which comes with it \cite{Landau1993, Helm2009, Helm2011, Crow2014, Trapani2017, Gregoratti2003}. Specifically, the pure dephasing decoherence, for principal systems with two-dimensional Hilbert space, and depolarizing noise, for all dimensionalities, has been simulated classically \cite{Crow2014}. 

It has been also shown that some kind of random unitary dynamics, producing classical noise, is capable of retrieving coherence, quantum correlations and entanglement between the parts of a composite quantum system, such as two qubits, that are locally interacting with the classical noise  
\cite{Franco2017, Orieux2015, DArrigo2014, Leggio2015, Trapani2015}.

Moreover, the effect of entanglement between a system and one of the main two types of quantum environments, the spin bath \cite{Prokofev2000, Tupitsyn97}, can be simulated by that of the other type, the oscillator bath \cite{Caldeira83,Leggett87}, in the weak- \cite{Caldeira83, Caldeira93, Prokofev2000, Weiss12, Schlosshauer2007} and strong-coupling \cite{Halataei2017mapping} limits of the spin bath, while the environments and their entanglement with the principal system are strikingly different in nature \cite{Prokofev2000, Prokofev93}. 

What is the role of entanglement in quantum decoherence? And what is the distinction between quantum and classical noises? This paper attempts to investigate these questions further. 

For a single qubit, it is well known that every doubly stochastic (or unital) channel can be represented as a random unitary channel \cite{Landau1993, Nielsen2010, Alicki2007, Andersson2007}. A doubly stochastic channel is a completely positive map on the Hilbert space of the principal system that maps the completely mixed state onto itself. A subclass of unital channels is made up of random unitary channels which are convex combinations of unitary transformations:
\beq
\m{E}(\rho) = \sum_i p_i U_i \rho U_i^\dagger \qquad \left( p_i >0, \sum_i p_i = 1 \right).
\eeq
We extend the idea of random unitary channels to random unitary expansions by letting the unitary operators $U_i$ depend on the initial state of the system-plus-environment. We show that for a single qubit with initial pure states $\rho(t_i)$ not only doubly stochastic operations but all quantum evolutions have random unitary expansions:
\beq \label{ruexp}
\rho(t) = \sum_\alpha p_\alpha U_\alpha \rho(t_i) U_\alpha^\dagger
\eeq
where $U_\alpha$ is a function of time and the initial state of the system-plus-environment. A Lebesgue integral over an infinitely uncountable set of index $\alpha$ is intended by the sum above.

For mixed initial state $\rho(t_i)$ of a single qubit, we show that a random unitary expansion is possible in the following sense:
\beq \label{rumixed}
\rho(t) = \sum_\alpha p_\alpha U_\alpha \rho_\alpha(t_i) U_\alpha^\dagger 
\eeq
where
\beq \label{rumixed0}
\rho(t_i) = \sum_\alpha p_\alpha \rho_\alpha (t_i)
\eeq
and $\rho_\alpha(t_i)$ are some pure states. 

We construct an explicit time-continuous classical model that simulates the effect of entanglement and derives the above results. For the sake of simplicity we first introduce the classical simulation of quantum models in which the system starts in pure states and the universe (system plus environment) starts in product states. We devote Secs. \ref{QM}--\ref{Simulation} to this case and describe the quantum and classical models and prove their equivalence. In Sec. \ref{MixedStates} we relax the initial-state assumption and let the system start in a mixed state and the universe in a non-product state. We show that a random unitary expansion is still possible in this case. Finally, in Sec. \ref{Ex} we give three examples for the case of initial pure states. The first example is the simulation of quantum recurrence in which the entropy decreases in the intermediate stage of evolution from an almost maximum value of $\ln 2$ to zero! The second example simulates pure dephasing decoherence. The last example is a simulation of amplitude damping, which could not be achieved in the previous classical models to this general extent \cite{Trapani2017, delCampo2017}.


\section{Quantum Model for Initial pure states} \label{QM}
In this section through Sec. \ref{Simulation} we consider all quantum models with the following four properties: (1) The central system {\it S} is a single qubit. (2) The qubit interacts with an arbitrary quantum bath {\it B}. (3) The initial state is a product state 
\beq \label{pr}
\rho_U(t_i) = \rho^Q (t_i) \otimes \rho_B(t_i)
\eeq
where $t_i$ is the initial time, $\rho_U$ is the density matrix of the universe (system plus bath), and $\rho^Q(t_i)$, $\rho_B(t_i)$ are the initial density matrices of the system and the bath (We use superscript $Q$ for the density matrix of the system, instead of subscript $S$, to emphasize that this density matrix is associated with the quantum model). (4) The system is initially in a pure state,
\beq
\rho^Q(t_i) = \ket{\Psi_i}\bra{\Psi_i}.
\eeq

The total Hamiltonian of the above quantum models can be decomposed into three parts as usual:  
\beq \label{HU}
H_U = H_S + H_{int} + H_B
\eeq
where $H_U$, $H_S$, $H_B$ and $H_{int}$ are the Hamiltonian of the universe, the qubit, the bath, and the interaction Hamiltonian respectively. The density matrix of the universe evolves by the evolution unitary operator 
\beq
U(t,t_i) = e^{-i H_U (t-t_i)}. 
\eeq
where we have set $\hbar = 1$. At each $t$ the density matrix of the universe is 
\beq \label{evolution}
\rho_U (t) = U(t,t_i) \rho_U(t_i) U(t,t_i)^{\dagger}.
\eeq
The quantity of interest here is the {\it reduced} density matrix of the system, which can be obtained by taking the trace of $\rho_U(t)$ over a basis of the bath
\beq 
\rho^Q(t) = \text{Tr}_{B} \left[ \rho_U(t) \right]
\eeq
Choosing some basis for the Hilbert space of the qubit, one can write $\rho^Q(t)$ in its matrix form
\beq \label{rhoQ}
\rho^Q(t) = 
\begin{pmatrix}
\rho^Q_{00}(t) & \rho^Q_{01}(t)  \\
\rho^Q_{10}(t)  & \rho^Q_{11}(t)
\end{pmatrix} 
\eeq
Since the evolution is quite arbitrary there are only a few general statements that one can make about $\rho^Q(t)$. Three of them are particularly useful in our discussion:
\beqn
\label{positive} & &\rho^Q_{00}(t), \rho^Q_{11}(t) \ge 0 \\
\label{traceunity} &&\rho^Q_{00}(t) + \rho^Q_{11}(t) = 1 \\
\label{offsmall} &&\abs{\rho^Q_{10} (t)} \le \sqrt{\rho^Q_{00} (t) \rho^Q_{11} (t) } .
\eeqn
The first two are well known. The third one can be derived from the positivity condition of the reduced density matrix, which implies $\det (\rho^Q) \ge 0$. The positivity of the reduced density matrix itself can be derived easily from the positivity of the universe density matrix. In App. \ref{app1} we give a proof for \eqref{offsmall}, which is finer than the positivity condition and may be used for generalization of the argument in higher dimensions. We shall use (\ref{positive})-(\ref{offsmall}) in constructing the classical model in the next section. 


\section{Classical Model for Initial pure states} \label{CM}
The classical model consists of a stochastic magnetic field, which acts on the qubit with Hamiltonian,
\beq
H_{Cl}(t) = \f{1}{2} \vec{B}(t) \cdot \vec{\sigma}
\eeq
where $\vec{B}(t) = (B_x(t), B_y(t), B_z(t))$ and $\vec{\sigma} = (\sigma_x, \sigma_y, \sigma_z)$ are Pauli matrices.  Each $B_j = (B_j (t): t_i \le t)$, for $j= x,y, z$, is a random process in the standard sense \cite{Hajek2015}. That is, each $B_j$ is a family of random variables $B_j(t)$ defined on a probability space $(\Omega, \m{F}, P)$ where $\Omega$ is the sample space, $\m{F}$ is a set of subsets of $\Omega$, and $P$ is a probability measure on $\m{F}$. By definition, for each $t$ fixed, the random variable $B_j(t)$ is a function from the sample space $\Omega$ to the real line: $\omega \mapsto B_j(t,\omega)$. Here $\omega$ are elements of $\Omega$. For each $\omega$ fixed, $\vec{B}(t,\omega)$ is a function of $t$, called the sample path (or noise history) corresponding to $\omega$, so $\omega$'s label the sample paths. 

On each sample path $\omega$, the qubit evolves from an initial state $\rho^{Cl} (t_i)$ to state $\rho^{Cl}_\omega (t)$ at each $t \ge t_i$. The initial state does not depend on $\omega$, however at any later time the state depends on $\omega$. The evolution operator for sample path $\omega$ is 
\beq \label{UClw}
U^{Cl}_\omega = \m{T} \exp[ - i \int_{t_i}^{t} H^{Cl}_\omega (t) dt ]
\eeq
where $\m{T}$ denotes time ordering and 
\beq \label{HCLw}
H^{Cl}_\omega(t) =  \f{1}{2} \vec{B}(t,\omega) \cdot \vec{\sigma}. 
\eeq
The evolved state at time $t$ can be written as 
\beq
\rho_\omega^{Cl} (t) = U^{Cl}_\omega \rho^{Cl} (t_i) U^{Cl \dagger}_\omega .
\eeq
In a classical noise model, the standard approach \cite{Crow2014} is to consider the density matrix of the qubit at time $t$ as the expectation value of  $\rho_\omega^{Cl} (t)$ over all sample paths, 
\beq
\label{finite} \rho^{Cl} (t) = \la \rho_\omega^{Cl} (t) \ra_\omega,
\eeq
where $\la \ra_\omega$ denotes the expectation function. 

Our goal is to construct the random magnetic field process $\vec{B}(t)$ such that $\rho^{Cl} (t) = \rho^Q (t)$ when the two models start from the same initial state  $\rho^{Cl} (t_i) = \rho^Q (t_i)$.

To this end, we begin with introducing, on a probability space $(\Omega, \m{F}, P)$,  a random process $\Phi = (\Phi(t) : t_i \le t )$ with the following properties: 
\begin{enumerate}
\item For all $\omega \in \Omega$, $\Phi(t_i, \omega) = 0$ 
\item For each $\omega$ fixed, $\Phi(t,\omega)$ is differentiable with respect to $t$. 

\item For each $t$ fixed, the probability density function of the random variable $\Phi(t)$ is a Gaussian with mean zero and variance $\sigma^2(t)$,
\beq \label{Gauss}
p_\Phi (\phi,t) = \f{1}{\sqrt{2 \pi \sigma^2(t)}} \exp \left[ - \f{\phi^2}{2 \sigma^2(t)} \right].
\eeq
where the probability density function $p_\Phi(\phi,t)$ for each $t$ is defined from the probability measure $P$ as 
\begin{align} \label{Gauss2}
p_\Phi(\phi, t) = \lim_{\epsilon \rightarrow 0} \f{1}{\epsilon} P\{ \omega: \phi \le \Phi(t,\omega) \le \phi + \epsilon \}
\end{align}
\item The variance $\sigma^2(t)$ is
\beq \label{sigma2}
\sigma^2(t) = - 2 \lim_{s \rightarrow t} \ln \f{\abs{\rho_{10}^Q (s)} }{\sqrt{\rho_{11}^Q(s) \rho_{00}^Q(s) } }
\eeq
\end{enumerate}
Inequality (\ref{offsmall}) guarantees that the right hand side of \eqref{sigma2} is nonnegative. Hence, it can be considered as the variance of a Gaussian distribution. In the case that $\sigma^2(t) = 0$ the Gaussian distribution becomes a delta function. For the case $\sigma^2(t)= \infty$ the distribution \eqref{Gauss} is interpreted as a uniform distribution over the entire real line. App. \ref{app2} gives a construction of $\Phi(t,\omega)$ which satisfies the above properties.

By use of the probability density function (\ref{Gauss}-\ref{Gauss2}), we can calculate the expectation of any function of $\Phi(t,\omega)$ for each $t$ fixed. As we shall see, it is particularly useful to calculate $\la \exp [\pm i \Phi(t,\omega)] \ra_\omega$:
\beqn
\nn \la e^{\pm i \Phi(t,\omega)} \ra_\omega &=& \int_{-\infty}^{\infty} e^{\pm i \phi} P\{ \omega: \phi \le \Phi(t,\omega) \le \phi + d\phi \} \\
\nn &=& \int_{-\infty}^\infty e^{\pm i \phi} \ p(\phi,t) \ d \phi = e^{-\sigma^2(t)/2} \\ \label{average}
&=& \f{\abs{\rho_{10}^Q (t)} }{\sqrt{\rho_{11}^Q(t) \rho_{00}^Q(t) } }
\eeqn
where the first integral above is a Lebesgue integral that is written in terms of an ordinary Reimann integral on the second line by use of the probability density function. 

We define for each $\omega$ fixed, functions $a(t,\omega)$, $b(t,\omega)$:
\beqn \label{a}
a(t,\omega) &=& \sqrt{\rho_{00}^Q(t)} \\ \label{bb}
b(t,\omega) &=& \sqrt{\rho_{11}^Q(t)} \ e^{ i \ \text{Arg}[\rho_{10}(t)]} e^{i \Phi(t,\omega)}
\eeqn  
Here Arg is the argument function over complex numbers (e.g. $z = \abs{z} e^{i \text{Arg}[z]}$). Note that $a(t,\omega)$ is a deterministic function of $t$ and is independent of sample path $\omega$, however, $b(t,\omega)$ is a random function and depends on the sample path. Nevertheless, on each sample path one always has the identity
\beq \label{equality}
\abs{a(t,\omega)}^2 + \abs{b(t,\omega)}^2 = \rho_{00}^Q(t) + \rho_{11}^Q(t) = 1  
\eeq
where we used Eq. \eqref{traceunity} in the last equality.

We are now ready to give the explicit form of $\vec{B}(t,\omega)$ :
\begin{align} \label{ep}
B_z(t,\omega) &= i [ \dot{a}(t,\omega) \ a(t,\omega)^* + \dot{b}(t,\omega)^* \ b(t,\omega) ] \\
B_+(t,\omega) &= - i [ \dot{a}(t,\omega)^* b(t,\omega) - \dot{b}(t,\omega) a(t,\omega)^* ] \\
B_x(t,\omega) &= \text{Re} \ B_+(t,\omega) \\ \label{By}
B_y(t,\omega) &= \text{Im} \ B_+(t,\omega) 
\end{align}
where overdots denote derivatives with respect to $t$. $B_x(t,\omega)$ and $B_y(t,\omega)$ are by definition real-valued. It is easy to show that $B_z(t,\omega)$ is also real-valued:
\beqn
\nn B_z(t,\omega) - B_z(t,\omega)^* &=& i [ \dot{a} a^* + \dot{a}^* \ a +  \dot{b}^* \ b + \dot{b} \ b^*] \\ \label{Bz}
&=& i \f{\p}{\p t} [a a^* + b b^*] = 0
\eeqn
Here we omitted $(t,\omega)$ dependencies of $a(t,\omega)$, $b(t,\omega)$ for brevity and used \eqref{equality} in the last step. Eq. \eqref{Bz} implies that $B_z(t,\omega)$ is real-valued. Thus (\ref{ep})-(\ref{By}) describe a well-defined stochastic magnetic field. Substituting them in Eq. \eqref{HCLw} one obtains the Hamiltonian $H^{Cl}_\omega (t)$ of the classical model on each sample path. 

We note that the classical Hamiltonian $H_\omega^{Cl}(t)$ depends on the total Hamiltonian of the quantum model $H_U$, the time elapsed from the beginning of the evolution and the initial state of the universe $\rho_U(t_i)$. This is because $a(t,\omega)$ and $b(t,\omega)$, which constitute $B(t,\omega)$, depend on $\rho^Q(t)$ which in turn depends on $\rho_U(t)$. The latter depends on $H_U$, $t$ and $\rho_U(t_i)$ (see Eqs. \eqref{HU}-\eqref{rhoQ}). Thus $H^{Cl}_\omega(t)$ not only is a function of the Hamiltonian of the quantum model, but also depends of the initial state of the universe. 


\section{Equivalence of the Quantum and Classical Models for initial pure states} \label{Simulation}
We assume that the qubit in the classical model starts from the same initial pure state as in the quantum model,
\beq
\rho^{Cl} (t_i) = \rho^Q (t_i) = \ket{\Psi_i}\bra{\Psi_i}.
\eeq
Since the initial state is a pure state, on each sample path $\omega$ the qubit evolves according to the time-dependent Schr\"{o}dinger equation 
\beq \label{sch}
i \ket{\dot{\Psi}(t,\omega)} = H^{Cl}_\omega(t) \ket{\Psi(t,\omega)}
\eeq
The solution of this Schr\"{o}dinger equation on each sample path is  
\beq \label{psi}
\ket{\Psi(t,\omega)} = \begin{pmatrix}
a(t,\omega) \\
b(t,\omega) 
\end{pmatrix}
\eeq
as we demonstrate below: Firstly, since $\Phi(t_i,\omega) = 0 $ for all sample paths, from definitions \eqref{a}-\eqref{bb} one can see that 
\beq
\ket{\Psi(t_i,\omega)} = \ket{\Psi_i}
\eeq
for each $\omega$, modulus an overall phase factor. Secondly, $\ket{\Psi(t,\omega)}$ of Eq. \eqref{psi} satisfies the Schr\"{o}dinger equation \eqref{sch}:
\begin{eqnarray}
\nn H_\omega^{Cl} \ket{\Psi(t,\omega)} &=& 
\begin{pmatrix}
B_z & B_+^* \\
B_+ & -B_z
\end{pmatrix} 
\begin{pmatrix}
a \\
b 
\end{pmatrix} \\ \nn
&=& \begin{pmatrix}
i \dot{a} [a^*a + b^* b] \\
i \dot{b} [a^*a + b^* b] 
\end{pmatrix} \\ \label{Scheq}
&=& \begin{pmatrix}
i \dot{a} \\
i \dot{b}  
\end{pmatrix} = i \ket{\dot{\Psi}(t,\omega)}
\end{eqnarray}
where we used \eqref{equality}. Thus, $\ket{\Psi(t,\omega)}$ is the solution of \eqref{sch}. 

The density matrix of the qubit on each sample path is then
\beq \label{rhoclw}
\rho^{Cl}_\omega(t) = \ket{\Psi(t,\omega)}\bra{\Psi(t,\omega)} = 
\begin{pmatrix}
\abs{a}^2 & a b^* \\
b a^* & \abs{b}^2
\end{pmatrix}
\eeq
The density matrix of the classical model is the average of the density matrices of all sample paths
\beq \label{rhocl}
\rho^{Cl}(t) = \la \rho^{Cl}(t) \ra_\omega = 
\begin{pmatrix}
\la \abs{a}^2 \ra_\omega & \la a b^* \ra_\omega \\
\la b a^* \ra_\omega & \la \abs{b}^2 \ra_\omega
\end{pmatrix}.
\eeq
Since $\abs{a(t,\omega)}^2 = \rho_{00}^Q(t)$ and $\abs{b(t,\omega)}^2 = \rho_{11}^Q(t)$ are deterministic functions, 
\beqn \label{a2}
\la \abs{a(t,\omega)}^2 \ra_\omega &=& \rho_{00}^Q(t) \\ \label{b2}
\la \abs{b(t,\omega)}^2 \ra_\omega &=& \rho_{11}^Q(t) .
\eeqn
For the off-diagonal term $\la b a^* \ra_\omega$ we have
\beq 
\la b(t,\omega) a(t,\omega)^* \ra_\omega = \sqrt{\rho_{00}(t) \rho_{11}(t)} e^{  i \ \text{Arg}[\rho_{10}(t)]} \la e^{i \Phi(t,\omega)} \ra_\omega
\eeq
By use of Eq. \eqref{average}, the above expression simplifies to 
\beq \label{ba}
\la b a^* \ra_\omega =  \abs{\rho_{10}^Q (t)} e^{  i \ \text{Arg}[\rho^Q_{10}(t)]} = \rho^Q_{10} (t)
\eeq
Similarly, one obtains
\beq \label{ab}
\la a b^* \ra_\omega =  \abs{\rho_{10}^Q (t)} e^{ - i \ \text{Arg}[\rho^Q_{10}(t)]} = \rho^Q_{01} (t).
\eeq
Finally, by substituting \eqref{a2}, \eqref{b2}, \eqref{ba}, and \eqref{ab} into \eqref{rhocl} we obtain
\beq \label{ClQ}
\rho^{Cl}(t) = 
\begin{pmatrix}
\rho^Q_{00}(t) & \rho^Q_{01}(t) \\
\rho^Q_{10}(t) & \rho^Q_{11}(t)
\end{pmatrix} = \rho^Q(t)
\eeq
just as desired. Hence, the classical model simulates the quantum model exactly. 

Summarizing, we built an stochastic magnetic field and hence a classical Hamiltonian for each history of noise. Then we showed that the density matrix of the qubit in this classical model at each moment of time is equal to the reduced density matrix of the quantum model at that time. 

One can write Eq. \eqref{ClQ} in a more familiar form  
\beq
\rho^Q(t) = \int_{-\infty}^{\infty} d \phi \ p(\phi,t) \ U_\omega^{Cl} \ket{\Psi_i} \bra{\Psi_i)} U_\omega^{Cl^\dagger} 
\eeq
where $p(\phi,t) d\phi = P\{\omega: \phi \le \Phi(t,\omega) \le \phi+ d \phi\}$ and $U^{Cl}_\omega = U^{Cl}_\omega(t; H_U; \rho_U(t_i))$. This form demonstrates that every quantum evolution of an open two dimensional system has a random unitary expansion (see Eq. \eqref{ruexp}).   

\section{Quantum and Classical Models for Initial Mixed States} \label{MixedStates}
The simulation of mixed states is similar to the one for pure states, apart from a few modifications that we mention below. For the quantum model, we relax conditions (3) and (4) of Sec. \ref{QM} for initial states of the universe and the principal system (a single qubit), and let them to start in any arbitrary states. That is the universe can start in a non-product state $\rho_U(t_i)$ and the system can start in a mixed state $\rho^Q(t_i) = \text{Tr}_B[\rho_U(t_i)]$. The universe evolves according to the unitary evolution of Eq. \eqref{evolution}. Relations \eqref{positive}-\eqref{offsmall} still hold since they do not depend on the initial states. Therefore we can use them in building the classical model as before. 

The classical model follows the classical model of Sec. \ref{CM} verbatim, except we relax the first property of the random phase $\Phi(t,\omega)$: It does not start from zero for all sample paths, rather it obeys the Gaussian distribution of Eq. \eqref{Gauss} for all times including the initial time $t_i$. Thus, $\Phi(t)$ is a differentiable random process with Guassian distribution whose mean and variance are zero and $\sigma^2(t)$ of Eq. \eqref{sigma2}, respectively. Since the initial state is a mixed state, $\sigma(t_i) \ne 0$. We construct $\Phi(t,\omega)$ as in App. \ref{app2}, Eq. \eqref{PhiZ}. Because $\sigma(t_i) \ne 0$, not all $\Phi(t_i, \omega)$ are equal to zero, as expected. The random magnetic field and the classical Hamiltonian follows Eqs. \eqref{ep}-\eqref{By} and \eqref{HCLw}, as before.

To prove the equivalence of the classical model and the quantum model in the case of initial mixed state we begin with constructing wave functions $\ket{\Psi(t,\omega)}$ as in Eq. \eqref{psi}. By use of Eqs. \eqref{rhoclw}-\eqref{ClQ} one can see that the density matrix of the quantum system at anytime, including the initial time $t_i$, can be expanded in terms of these pure states:
\beq \label{rhoQt}
\rho^Q(t) = \int_{-\infty}^{\infty} d\phi \ p(\phi,t) \ \ket{\Psi(t,\omega)} \bra{\Psi(t,\omega)} 
\eeq
One notes that wave functions $\ket{\Psi(t,\omega)}$ now do not start from the same value at $t=t_i$ because for two different sample paths $\omega$, $\omega'$, the initial value of the random phase can be different $\Phi(t_i,\omega) \ne \Phi(t_i,\omega')$. Nevertheless, $\ket{\Psi(t,\omega)}$ satisfy Schr\"{o}dinger equation \eqref{sch} on each sample path, as shown in Eq. \eqref{Scheq}. Therefore, one can obtain $\ket{\Psi(t,\omega)}$ by evolving $\ket{\Psi(t_i,\omega)}$ through $U^{Cl}_\omega$:
\beq \label{Psitomega}
\ket{\Psi(t,\omega)} = U^{Cl}_\omega \ket{\Psi(t_i,\omega)}
\eeq
where $U^{Cl}_\omega$ is defined in Eq. \eqref{UClw}. Substituting \eqref{Psitomega} into Eq. \eqref{rhoQt} we obtain
\beq \label{rhoQt2}
\rho^Q(t) = \int_{-\infty}^{\infty} d\phi \ p(\phi,t) \ U^{Cl}_\omega \ket{\Psi(t_i,\omega)} \bra{\Psi(t_i,\omega)} U^{Cl^\dagger}_\omega.
\eeq
At $t=t_i$, $U^{Cl}_\omega$, which is a function of time, is the identity operator and Eq. \eqref{rhoQt2} reduces to 
\beq \label{rhoQt3}
\rho^Q(t_i) = \int_{-\infty}^{\infty} d\phi \ p(\phi,t) \ \ket{\Psi(t_i,\omega)} \bra{\Psi(t_i,\omega)} .
\eeq
Eqs. \eqref{rhoQt2}-\eqref{rhoQt3} are the random unitary expansions we intended to find for arbitrary evolution of a single qubit with initial mixed state (see Eqs. \eqref{rumixed}-\eqref{rumixed0}). In the sense of these equations the classical model and the quantum model are equivalent for initial mixed states as well.


\section{Examples} \label{Ex}
We consider three examples for classical simulation of quantum models. In all the examples we assume that the universe starts in the product state $\rho_U(0) = \ket{\Psi_i} \bra{\Psi_i} \otimes  \rho_B(0)$ where $\ket{\Psi_i} =  \alpha \ket{0} + \beta \ket{1}$ and $\rho_B(0)$ will be specified for each example. We determine $\sigma^2(t)$, $a(t,\omega)$ and $b(t,\omega)$ in each example. The stochastic magnetic field $B(t,\omega)$ and the classical Hamiltonian $H_{Cl}(t,\omega)$ can then be constructed by Eqs. \eqref{ep}-\eqref{By}, \eqref{HCLw} and the discussion of Sec. \ref{CM}. 

\subsection{Quantum recurrence} 
Consider a spin-boson Hamiltonian at zero temperature 
\beq \label{SB}
H_U = \f{1}{2} \omega_0 \sigma_z + \sigma_z \sum_{n=1}^N (g_n a_n^\dagger + g_n^* a_n) + \sum_{n=1}^N \omega_n a_n^\dagger a_n
\eeq
where $N$ is finite and the frequencies of the bath are commensurable (i.e. for each $\omega_n$, $\omega_m$ there are integer numbers $p_n$ , $p_m$ such that $\omega_n /\omega_m = p_n/p_m$). The bath is initially in its ground state. The evolution of the reduced density matrix is then \cite{Schlosshauer2007} 
\beq \label{examp1}
\rho^{Q}(t) = 
\begin{pmatrix}
\abs{\alpha}^2 & \alpha \beta^* \ e^{-i \omega_0 t - \Gamma(t)} \\
\alpha^* \beta \ e^{i \omega_0 t - \Gamma(t)} & \abs{\beta}^2
\end{pmatrix} 
\eeq
where
\beq
\Gamma(t) =  \sum_{n=1}^N 4 \f{\abs{g_n}^2}{\omega_n^2} (1 - \cos \omega_n t)
\eeq
Since $\omega_n$'s are commensurable, $\Gamma(t)$ is a periodic function. It starts at $\Gamma(0) = 0$
 and returns to zero with some period $P$. Between two nodes of function $\Gamma(t)$, however, the value of the function can be large if $N$ is large or coupling constants $g_n$ are significant. For example, for $\abs{g_n} = \omega_n = 2 \pi n / P$ and $N= 30$ the average value of the function between two nodes is $\Gamma(t) \approx 120$. This gives rise to decoherence factor $\exp [-\Gamma(t)] \approx 10^{-53}$ in the off diagonal elements of \eqref{examp1}. This implies that for $\alpha = \beta = 1/\sqrt{2}$ the system that started in a pure state with entropy $S[\rho^Q(0)] = Tr [\rho^Q(0) \ln \rho^Q(0)] = 0$ evolves to nearly the completely maxed state
\beq \label{examp2}
\rho^{Q}(t) \approx 
\begin{pmatrix}
\f{1}{2} & 0 \\
0 & \f{1}{2}
\end{pmatrix}  \qquad 0 \ll t \ll P
\eeq
with maximum possible entropy $S[\rho^Q(t)] \approx \ln 2$ for most of the times between, for example, $t= 0$ and $t=P$, and then the entropy of the system {\it decreases} and the system returns to the original pure state at $t=P$ with $S[\rho^Q(P)] = 0$. 

In terms of Bloch vector the above process describes a contraction of the Bloch sphere to {\it almost} a point and then an expansion of it to its full size.
 
Although the entropy decreases in the intermediate stage in the above process, it can still be simulated classically. The variance $\sigma(t)^2$ defined in \eqref{sigma2} is in this case
\beq \label{sigma2G}
\sigma^2(t) = 2 \ \Gamma(t)
\eeq
and the functions $a(t,\omega)$, $b(t,\omega)$ of Eqs. (\ref{a})-(\ref{bb}) are
\beq \label{ab2}
a(t,\omega) = \abs{\alpha}, \qquad b(t,\omega) = \abs{\beta} \ e^{i \text{Arg}[\alpha^* \beta]} \ e^{i \omega_0 t} e^{i \Phi(t,\omega)}
\eeq
where $\Phi(t,\omega)$ is the random phase defined in points 1-4 of Sec. \ref{CM}.
\subsection{Pure dephasing}
One can also use Hamiltonian \eqref{SB} as an example of phase damping in the limit $N \rightarrow \infty$. Suppose the bath is in thermal equilibrium at some nonzero temperature $T$ and the spectral density function of the bath is ohmic: $J(\omega) = \sum_n \abs{g_n}^2  \delta(\omega - \omega_n) = 4^{-1} J_0  \omega  e^{-\omega/ \Lambda}$, where $J_0$ is a dimensionless constant and $\Lambda$ is a cut-off frequency. The evolution of the reduced density matrix can be described by Eq. \eqref{examp1} when one substitutes for $\Gamma(t)$ the following expression \cite{Schlosshauer2007}
\begin{align}
\nn \Gamma(t) &= \int_0^\infty d \omega \f{4 J(\omega)}{\omega^2} (1 - \cos \omega t) \coth (\omega/2 k_B T) \\ \label{Gt}
&= \f{J_0}{2} \ln (1 + \Lambda^2 t^2) + J_0 \ln \left[ \f{\sinh (\pi k_B T t)}{\pi k_B T t} \right]
\end{align}
Here $\Gamma(t)$ is an increasing function of time, which gives rise to the increase of the entropy of the system over time. There is no revival of coherence in this limit. The classical model is similar to the one in the previous subsection and is described  by Eqs. \eqref{sigma2G}-\eqref{ab2} where $\Gamma(t)$ is given by Eq. \eqref{Gt}.

\subsection{Amplitude damping}
Finally, consider an amplitude damping channel \cite{Nielsen2010}
\beq
\rho(t) = \begin{pmatrix}
1 - (1-\gamma(t))(1-\abs{\alpha}^2) & \alpha \beta^* \sqrt{1 - \gamma(t)} \\
\alpha^* \beta \sqrt{1 - \gamma(t)} & \abs{\beta}^2 (1-\gamma(t))
\end{pmatrix} 
\eeq
where the environment starts in the ground state and $\gamma(t)$ is the probability of decay of the qubit from the excited state to its ground state. For real physical processes, $\gamma(t)$ can be replaced by $(1 - e^{-t/T_1})$, where $T_1$ is the longitudinal relaxation time constant. 

An amplitude damping channel is not a unital channel and a general classical simulation of it has not been achieved in the literature, to our knowledge. The classical model of Sec. \ref{CM} gives such a simulation by letting the unitary operators of the classical model to depend on the initial state of the universe. The ingredients of the model are as follows 
\beqn
a(t,\omega) &=& \left[1 - (1-\gamma(t))(1-\abs{\alpha}^2)\right]^{1/2} \\
b(t,\omega) &=& \abs{\beta} \sqrt{1 - \gamma(t)} \ e^{i \text{Arg}[\alpha^* \beta]} \ e^{i \Phi(t,\omega)} \\
\sigma^2(t) &=& \ln \left(\abs{\alpha}^{-2} - (1-\gamma(t))(\abs{\alpha}^{-2}-1)\right).
\eeqn

\section{Conclusion}
In conclusion, we have constructed, for arbitrary quantum noises, a classical simulation of single-qubit models. We showed how entanglement between a qubit and an external bath can be modeled classically without using the bath. This was made possible by allowing the unitary operators in the classical model to depend on the initial state of the system and the bath. 

We demonstrated that the reduced density matrices of quantum models that start from initial pure states have random unitary expansions. For the quantum models that start from mixed states (and even non-product states of the system and bath) we showed that the density matrices can be expressed as a random unitary expansion of some pure states. 

The classical model was based on utilizing a differentiable random phase that has Gaussian distribution with time-varying variance. We gave the explicit expression for the stochastic magnetic field of the classical Hamiltonian. The field depends on the Hamiltonian of the quantum model, the time elapsed from the beginning of the evolution, and the initial state of the system and bath in the quantum model.   

Simulation of quantum dissipation such as amplitude damping had not been achieved in preceding classical models, except for short times and high temperatures. Here, we offered exact results for a general simulation of such a process (amplitude damping), for arbitrary long times, as well as of quantum recurrence and pure dephasing.

Entanglement with an external environment plays an important role in quantum dissipation and decoherence of open quantum systems, beyond doubt. However, the result of this paper and its preceding counterparts show that, as far as the simulation is concerned, the distinction between quantum and classical noises may not be apparent in systems with low dimensionality.  


\section*{Acknowledgement}
I would like to thank Mahdieh Piranaghl, Anthony J. Leggett (my PhD adviser), Bruce Hajek, Ehsan Shafiee and Richard Sawers for fruitful discussions, valuable hints and support.  
\begin{appendices}

\section{\\A property of the reduced density matrix} \label{app1}
In this appendix we prove the inequality \eqref{offsmall}, $\abs{\rho_{10}} \le \sqrt{\rho_{00} \rho_{11} }$, for any $2 \times 2$ reduced matrix. 
 
Quite generally, the density matrix of the universe can be written in terms of a statistical mixture of pure states of the universe, $\rho_U = \sum_n p_n \ket{\Psi_n} \bra{\Psi_n}$ where $\ket{\Psi_n}$ are pure states of the universe and $p_n$ are their statistical probabilities ($p_n \ge 0$, $\sum_n p_n = 1$.) 

One can choose basis $\{ \ket{i,\mu} \}$ for the universe which is a direct product of a basis of the system $\{ \ket{i} \}$, where $i = 0,1$,  and a basis of the bath $\{ \ket{\mu} \}$. Then one can expand the pure states $\ket{\Psi_n}$ in terms of this basis, $\ket{\Psi_n} = \sum_{i, \mu} c_{i,\mu}^n \ket{i,\mu}$ and rewrite the universe density matrix as 
\beq
\rho_U = \sum_{n; i,j; \mu, \nu}  p_n \ c_{i,\mu}^n \ c_{j,\nu}^{n^*} \ \ket{i,\mu} \bra{j,\nu}
\eeq
Now the reduced density matrix elements are as follows, 
\beqn
\rho_{00} &=& \sum_{n; \mu} p_n \ \abs{c_{0,\mu}^n}^2 \\
\rho_{11} &=& \sum_{n; \mu} p_n \ \abs{c_{1,\mu}^n}^2 \\
\rho_{10} &=& \sum_{n; \mu} p_n \ c_{1,\mu}^n \ c_{0,\mu}^{n^*}. 
\eeqn
$\abs{\rho_{01}}$ satisfies the following inequality
\beq \label{ineq}
\abs{\rho_{01}} = \abs{\sum_{n; \mu} p_n \ c_{1,\mu}^n \ c_{0,\mu}^{n^*} } \le \sum_{n; \mu} p_n \ \abs{c_{1,\mu}^n} \ \abs{c_{0,\mu}^{n^*}} .
\eeq
We define vectors $\vec{R_i}$
\beq
\vec{R}_i = (\sqrt{p_0} \ \abs{c_{i,0}^0}, \sqrt{p_0} \abs{c_{i,1}^0}, \cdots ; \sqrt{p_1} \ \abs{c_{i,0}^1}, \sqrt{p_1} \abs{c_{i,1}^1}, \cdots )
\eeq
Then we observe that $\rho_{00} = \abs{\vec{R}_0}^2$, $\rho_{11} = \abs{\vec{R}_1}^2$, and, from inequality \eqref{ineq}, 
\beq
\abs{\rho_{01}} \le \vec{R}_0 . \vec{R}_1.  
\eeq
Since, $\vec{R}_0 . \vec{R}_1 \le \abs{\vec{R}_0} \abs{\vec{R}_1}$ we conclude that 
\beq
\abs{\rho_{10}} \le \sqrt{\rho_{00} \rho_{11} }.
\eeq

\section{\\Construction of $\Phi(t,\omega)$} \label{app2}
$\Phi$ is a random process with Gaussian distribution whose variance $\sigma^2(t)$ is time dependent. We demanded that the process be differentiable with respect to time on each sample path in order to have a well-defined magnetic field in Eqs. (\ref{ep})-(\ref{By}). Such a process exists, as long as $\sigma^2(t)$ is differentiable with respect to time \cite{Hajekp}. There are many constructions for $\Phi$, depending on what correlation function one would like for the random process. The simplest construction is the following \cite{Hajekp}: Let $Z$ be a standard Gaussian random variable with mean zero and variance equal to unity. By definition of random variables, $Z$ is a function from a sample space $\Omega$ to the real line : $\omega \mapsto Z(\omega)$. Now let
\beq \label{PhiZ}
\Phi(t,\omega) = \sigma(t) Z(\omega),
\eeq
where $\sigma(t)$ is taken from Eq. (\ref{sigma2}). $\Phi(t,\omega)$ has the properties we wanted. For each $t$ fixed, the distribution of the process is Gaussian because the distribution of $Z$ is Gaussian. Also the mean and variance are
\begin{align}
& \la \Phi(t,\omega) \ra_\omega = \la \sigma(t) Z(\omega) \ra_\omega = \sigma(t) \la Z(\omega) \ra_\omega = 0 \\
& \la \Phi(t,\omega)^2 \ra_\omega = \sigma^2(t) \la Z(\omega)^2 \ra_\omega = \sigma^2(t)
\end{align}
as expected. For each $\omega$ fixed, $\Phi(t,\omega)$ is differentiable with respect to $t$ since $\sigma(t)$ is so. Finally, $\Phi(t_i,\omega) = 0$ for initial pure states because $\sigma(t_i) = 0$ for such states. In this case the right hand side of Eq. (\ref{sigma2}) is zero. For mixed states the right hand side of Eq. (\ref{sigma2}) is nonzero and $\sigma(t_i) \ne 0$. However, $\Phi(t_i,\omega)$ is not required in the classical model of Sec. \ref{MixedStates} to be zero either. Thus the random phase \eqref{PhiZ} works well for this case, too. 

\end{appendices}

\bibliographystyle{unsrt}
\bibliography{thesisrefs}

\end{document}